\documentclass[acmsmall]{acmart}
\usepackage{enumerate}
\usepackage{algorithm2e}
\RestyleAlgo{ruled}
\usepackage{multirow}
\usepackage{tcolorbox}
\newtcolorbox{findingbox}[2][]{colback=black!5!white,colframe=white!20!black,fonttitle=\bfseries, title=#2,#1}

\usepackage{courier} 
\usepackage{listings, xcolor}

\definecolor{javapurple}{rgb}{0.5,0,0.35} 
\definecolor{linenumbergray}{rgb}{0.5,0.5,0.5}
\lstdefinestyle{Java-github}{
        basicstyle=\ttfamily\tiny,
        language=Java,
        commentstyle=\color{linenumbergray},
        stringstyle=\color{javapurple},
        keywordstyle=\color{red},
        morekeywords={@Test},
        morecomment=[s][\color{linenumbergray}]{/**}{*/},
        numbers=left,
        numberstyle=\tiny\color{linenumbergray},
        numbersep=2.5pt,
        xleftmargin=1em,
        moredelim=**[is][\color{javapurple}]{@h@}{@h@},
        morecomment=[f][{\btHL[fill=gitdel]}]-,
        morecomment=[f][{\btHL[fill=gitadd]}]+,
        breaklines = true,
}

\lstdefinestyle{prompt}{
        basicstyle=\ttfamily\tiny,
        language=Html,
        commentstyle=\color{linenumbergray},
        stringstyle=\color{javapurple},
        keywordstyle=\color{red},
        morekeywords={@Test},
        morecomment=[s][\color{linenumbergray}]{/**}{*/},
        numbers=left,
        numberstyle=\tiny\color{linenumbergray},
        numbersep=2.5pt,
        xleftmargin=1em,
        moredelim=**[is][\color{javapurple}]{@h@}{@h@},
        morecomment=[f][{\btHL[fill=gitdel]}]-,
        morecomment=[f][{\btHL[fill=gitadd]}]+,
        breaklines = true,
}
\AtBeginDocument{%
  }

\setcopyright{acmlicensed}
\copyrightyear{2024}
\acmYear{2024}
\acmDOI{XXXXXXX.XXXXXXX}

\acmJournal{JACM}
\acmVolume{37}
\acmNumber{4}
\acmArticle{111}
\acmMonth{8}

\newcommand{\TitleArxiv}{Generating executable oracles to check conformance of client code to requirements of JDK Javadocs using LLMs}

\begin{document}

\title{\TitleArxiv}

\author{Shan Jiang}
\affiliation{%
  \institution{The University of Texas at Austin}
  \city{Austin}
  \country{USA}}
\email{shanjiang@utexas.edu}

\author{Chenguang Zhu}
\affiliation{%
  \institution{The University of Texas at Austin}
  \city{Austin}
  \country{USA}}
\email{cgzhu@utexas.edu}

\author{Sarfraz Khurshid}
\affiliation{%
  \institution{The University of Texas at Austin}
  \city{Austin}
  \state{Texas}
  \country{USA}}
\email{khurshid@ece.utexas.edu}

\renewcommand{\shortauthors}{Shan et al.}

\newcommand{\Comment}[1]{}

\begin{abstract}
  \Comment{
  Software testing is vital in the software development lifecycle. In modern development, the gap between natural language specifications and source code can introduce ambiguities, reducing testing efficiency. Test oracles are essential in detecting discrepancies between intended and actual functionality. While various test oracle generation techniques exist, they face limitations due to strict specification formats and difficulty in generalizing across test cases. These methods typically focus on verifying alignment between implementations and their project-specific documentation. However, there remains a gap in research on verifying if implementations align with broader language contracts, such as JDK Javadocs.
 
  Recently, large language models (LLMs) have been increasingly applied to various software engineering tasks, showcasing remarkable accuracy in capturing natural language information. In this paper, we introduce a novel method that leverages LLMs to generate test oracles from JDK Javadocs, enabling verification of whether software implementations adhere to JDK contracts. Our approach is motivated by the high quality and richness of information in JDK Javadocs, which we can effectively transform into test oracles.
  }
Software testing remains the most widely used methodology for
validating quality of code.  However, effectiveness of testing
critically depends on the quality of test suites used.  Test cases in
a test suite consist of two fundamental parts: (1) input values for
the code under test, and (2) correct checks for the outputs it
produces.  These checks are commonly written as assertions, and termed
test oracles.  The last couple of decades have seen much progress in
automated test input generation, e.g., using fuzzing and symbolic
execution.  However, automating test oracles remains a relatively less
explored problem area.  Indeed, a test oracle by its nature requires
knowledge of expected behavior, which may only be known to the
developer and may not not exist in a formal language that supports
automated reasoning.

Our focus in this paper is automation of test oracles for clients of
widely used Java libraries, e.g., \textit{java.lang} and \textit{java.util} packages.
Our key insight is that Javadocs that provide a rich source of
information can enable
automated generation of test oracles. Javadocs of the core Java
libraries are fairly detailed documents that contain natural language
descriptions of not only how the libraries behave but also how the
clients must (not) use them.  We use large
language models as an enabling technology to embody our insight into a framework for test oracle
automation, and evaluate it experimentally.
  Our experiments demonstrate that LLMs can generate oracles for checking normal and exceptional behaviors from Javadocs, with 98.8\% of these oracles being compilable and 96.4\% accurately reflecting intended properties.  Even for the few incorrect oracles, errors are minor and can be easily corrected with the help of additional comment information generated by the LLMs.
\end{abstract}



\maketitle

\section{Introduction}
Software testing is the most widely used methodology for validating code quality\cite{wang2024software}.  However, effectiveness of testing critically depends on the quality of test suites used.  A test suite consists of multiple test cases and each test case has two fundamental parts: (1) input values used to run the code under test, and (2) correctness checks run to validate the outputs it produces.  These checks are commonly written as assertions, and serve as executable
test oracles.  In the last few decades, researchers have made much progress in automated test input generation, e.g., using fuzzing\cite{zhu2022fuzzing,xia2023universal} and symbolic execution\cite{baldoni2018survey,khurshid2003generalized,king1976symbolic}. Generating test oracles is inherently challenging as it requires an understanding of the expected behavior, which often is known only to the developer and may not be documented in a formal language specification to automated reasoning. This problem occurs in different scenarios, including mobile applications, embedded systems, quantum systems, etc.\cite{meng2022batmapper,jeon2002embedding,lewis2023formal}. Besides, existing test oracle generation techniques typically focus on verifying alignment between implementations and their project-specific documentation\cite{liu2020automatic,toga,hossain2024togll}. However, there remains a gap in research on verifying if implementations align with broader language contracts, such as JDK Javadocs.


\begin{figure}
    \centering
    \includegraphics[width=.8\linewidth]{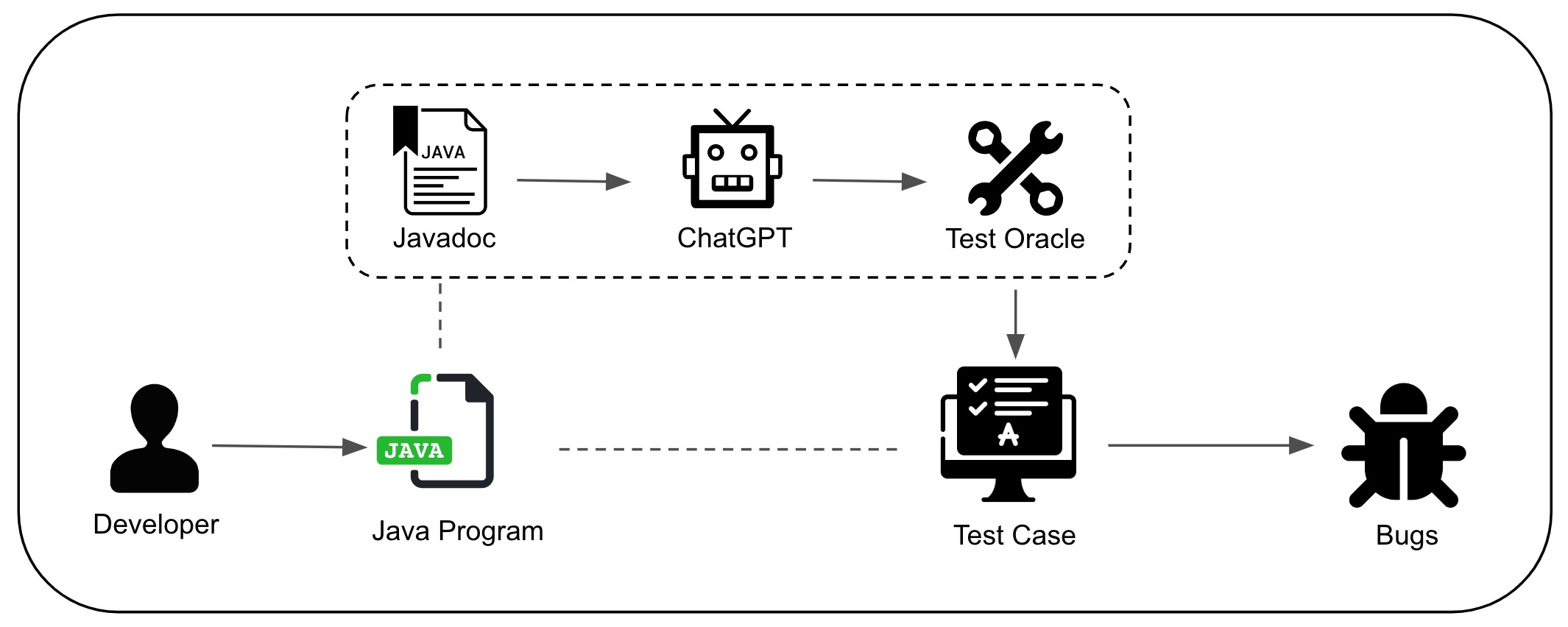}
    \caption{An overview of software testing workflow, the dashed box represents our work's scope}
    \label{workflow}
\end{figure}

Software bugs, ranging from unexpected exceptions to incorrect outputs, frequently stem from mismatches between the intended behavior described in natural language and its actual implementation in code \cite{afshan2013evolving,harman2010optimizing}. This problem has intensified with the rise of LLM-assisted programming, where developers use large language models (LLMs) to generate code from natural language prompts \cite{endres2024can,ouyang2023llm}. In particular, when working with Javadoc, client developers might unintentionally create inconsistencies by overlooking or misinterpreting constraints specified in the documentation. Therefore, a testing approach is essential to identify discrepancies between Javadoc descriptions and the corresponding client code that either overrides standard Java classes or utilizes standard Java interfaces.

 Test oracles play a crucial role in identifying discrepancies by ensuring that software behavior aligns with its intended functionality \cite{leitner2007contract,shin2024assessing}. Traditional specification mining techniques have largely focused on deriving oracles from structured code or constrained specifications \cite{tan2012tcomment}, often neglecting the rich potential of natural language descriptions. In contrast, while neural generative models \cite{toga} offer a more flexible approach, they face challenges in generating precise oracles due to the vast range of possible assertions. Furthermore, their effectiveness can vary significantly across different real-world Java projects, making it difficult to develop universally applicable test oracles \cite{hossain2023neural}. Ideally, a technique should accurately generate oracles that capture the developer's intent, maintain a low false positive rate, and be generalizable across projects to effectively test discrepancies between Javadoc and its client implementations.

In this paper, we explore the potential of LLMs as intermediaries between natural language documentation and Java test oracles. Recently, LLMs have shown strong capabilities in tasks that require both natural language comprehension and code-related functionalities \cite{li2023cctest,nam2024using}. Trained on extensive corpora, including large codebases and documentation, these models possess an implicit understanding of programming syntax and semantics. Research suggests that LLMs can effectively align user inputs with their pre-existing knowledge \cite{liu2024llm}, and they even display emergent abilities to handle tasks uncommon in their training data \cite{schaeffer2024emergent}. In our context, the richness of JDK Javadocs allows us to derive general contracts directly from documentation, without needing code implementation. These contracts, when formatted as oracles, provide essential guidelines for client developers to follow.


We introduce a novel methodology that utilizes LLMs to automate test oracle generation directly from JDK Javadocs, aiming to address subtle conventions that Java projects must conform to within JDK contracts. As illustrated in Figure \ref{workflow}, our approach integrates into the broader software testing process without requiring any actual code implementation. Unlike traditional methods that generate test oracles from each project's documentation for use within that project, our approach leverages JDK Javadocs as the primary input. We focus on the JDK because it is foundational to all Java applications, making its conventions essential for Java projects, and its documentation is meticulously maintained to ensure quality \cite{tan2008empirical,lee2012towards}. By extracting key properties from the JDK, our generated test oracles represent general contracts that all client Java projects should adhere to.

Our experimental results highlight the effectiveness of LLMs in generating relevant and accurate test oracles. Specifically, our approach successfully produces 97\% compilable test oracles without the need for additional compilation fix tools. Our technique also shows strong applicability and completeness: LLMs cover 90.3\% of the properties outlined in JDK Javadocs, with 96.0\% of the generated oracles accurately validating these properties. Additionally, we generate oracles for 98.9\% of exceptions defined in JDK Javadocs, with 97.2\% of these oracles effectively capturing the intended exceptions, demonstrating LLMs' capability in creating comprehensive and precise exception oracles. With high-quality comments and clear naming conventions, the generated oracles are intuitive and straightforward, facilitating efficient testing. These promising results underscore the potential of LLMs in identifying misalignments between software specifications and their implementations.


To summarize, this paper makes the following contributions:

\begin{itemize} 
\item \textbf{Idea.} We introduce the idea of test oracle automation using Javadocs for standard Java libraries in order to automate testing of the clients of these libraries.

\item \textbf{Approach.} We use large language models (LLMs) as an enabling technology to embody our insight into a technique for automating test oracles. We propose a prompt template which is flexible and allows generation of test oracles for a variety of JDK Javadocs.

\item \textbf{Evaluation.} We conducted an experimental evaluation using some of the most widely used JDK Java classes and interfaces from \textit{java.lang} and \textit{java.util} (e.g., \textit{Object}, \textit{String}, \textit{Map}, \textit{Set}, \textit{List}) and present the resulting data. Our findings confirm that LLMs can generate highly effective and applicable test oracles, capable of accurately capturing expected behaviors and intended exceptions within the JDK.

\item \textbf{Artifact.} We release our prompt template and the complete set of generated test oracles for reproducibility.  This allows developers to adapt our approach to their specific needs, ensuring that Java projects conform to JDK contract with minimal manual intervention.
\end{itemize}

\section{An Illustrative Example}

Inheritance promotes code reusability and modularity by allowing a subclass to inherit attributes and methods from a superclass. However, it also introduces challenges for software maintenance and testing \cite{kim2024exploring}. In the context of the Java Development Kit, failing to adhere to a superclass’s specifications can lead subclasses to violate JDK contracts, potentially undermining software integrity and reliability. To illustrate this, we use an example with two Java classes: \textit{Point} and \textit{Point3D}, highlighting a discrepancy caused by a subclass violating JDK contracts.

\begin{figure}[!t]
	\centering
	\begin{minipage}[t]{0.44\linewidth}
		\begin{lstlisting}[style = Java-github]
public class Point {
    private int x;
    private int y;

    public Point(int x, int y) {
        this.x = x;
        this.y = y;
    }

    public boolean equals(Object o) {
        if (! (o instanceof Point)) { return false; }
        Point p = (Point) o;
        return (this.x == p.x) && (this.y == p.y);
    }
}
            \end{lstlisting}
            \caption{\textit{Point} Class}
            \label{point}
	\end{minipage}
 \hspace{.3in}
	\begin{minipage}[t]{0.44\linewidth}
		\begin{lstlisting}[style = Java-github]
public class Point3D extends Point{
    private int z;

    public Point3D(int x, int y, int z) {
        super(x, y);
        this.z = z;
    }

    public boolean equals(Object o) {
        if (! (o instanceof Point3D)) { return false; }
        Point3D p = (Point3D) o;
        return super.equals(p) && (z == p.z);
    }
}
            \end{lstlisting}
            \caption{\textit{Point3D} Class}
		\label{point3D}
	\end{minipage}
 
\end{figure}

Figure \ref{point} shows the implementation of \textit{Point} class. It represents a point in two-dimensional space, with attributes for the \textit{x} and \textit{y} coordinates. The equals method in this class is overridden to check equality based on these coordinates. After verifying that the object being compared is an instance of the Point class, it compares the \textit{x} and \textit{y} coordinates for equality. 

Figure \ref{point3D} is the implementation of \textit{Point3D} class, which is a subclass of \textit{Point}. The \textit{Point3D} class extends the \textit{Point} to three-dimensional space by adding a z-coordinate attribute. Overriding the equals method in \textit{Point3D} requires calling the equals method of its superclass \textit{Point} to compare the \textit{x} and \textit{y} coordinates, and additionally comparing the z-coordinate.

\begin{figure}
    \centering
    \includegraphics[width=\linewidth]{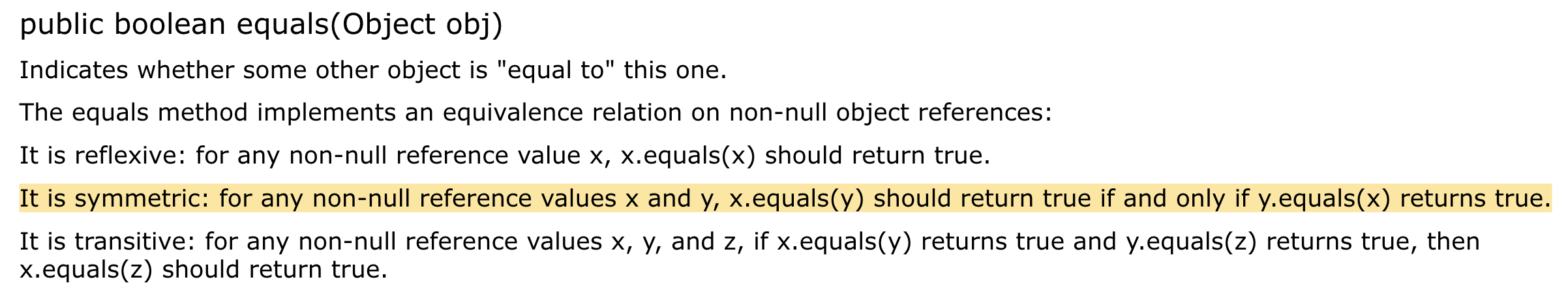}
    \caption{Symmetric property in the specification of \textit{java.lang.Object equals(Object o)} method}
    \label{sym}
\end{figure}

According to the specification of the \textit{java.lang.Object} class, which sits at the root of the Java class hierarchy, all overridden \textit{equals} methods must satisfy the \textbf{symmetric} property, as shown in Figure \ref{sym}. This property requires that for any non-null reference values \textit{x} and \textit{y}, \textit{x.equals(y)} should return true if and only if \textit{y.equals(x)} also returns true. However, the implementation of \textit{Point} and \textit{Point3D} violates this contract.


\begin{figure}
	\centering
	\begin{minipage}[t]{0.44\linewidth}
		\begin{lstlisting}[style = Java-github]
public void testSymmetric(){
  Point p = new Point(3, 4);
  Point3D p3 = new Point3D(3, 4, 5);
    assertTrue(p.equals(p3) == p3.equals(p));
}
            \end{lstlisting}
	\end{minipage}
 \hspace{.3in}
	\begin{minipage}[t]{0.44\linewidth}
		\begin{lstlisting}[style = Java-github]
boolean checkSymmetric(Object x, Object y) {
    if (x == null || y == null) return x == y;
    return x.equals(y) == y.equals(x);
}

            \end{lstlisting}
	\end{minipage}

 \caption{A test case revealing the discrepancy (left) and an oracle method for checking symmetric property (right)}
		\label{test}
 
\end{figure}

The test cases in Figure \ref{test} reveal a bug where the symmetry is violated. In this test case, while \textit{p3.equals(p)} returns false (as \textit{p3} considers the z-coordinate), \textit{p.equals(p3)} return true (as \textit{Point} only checks \textit{x} and \textit{y} coordinates, ignoring the additional z-coordinate in \textit{Point3D}).

This discrepancy breaches the symmetry requirement, highlighting a common oversight in software development where developers may unintentionally violate inherited behaviors. In more complex industry settings, the logic often becomes increasingly intricate, leading to more critical errors. 


\section{Test Oracle Generation}
\begin{figure}
    \centering
    \includegraphics[width=.9\linewidth]{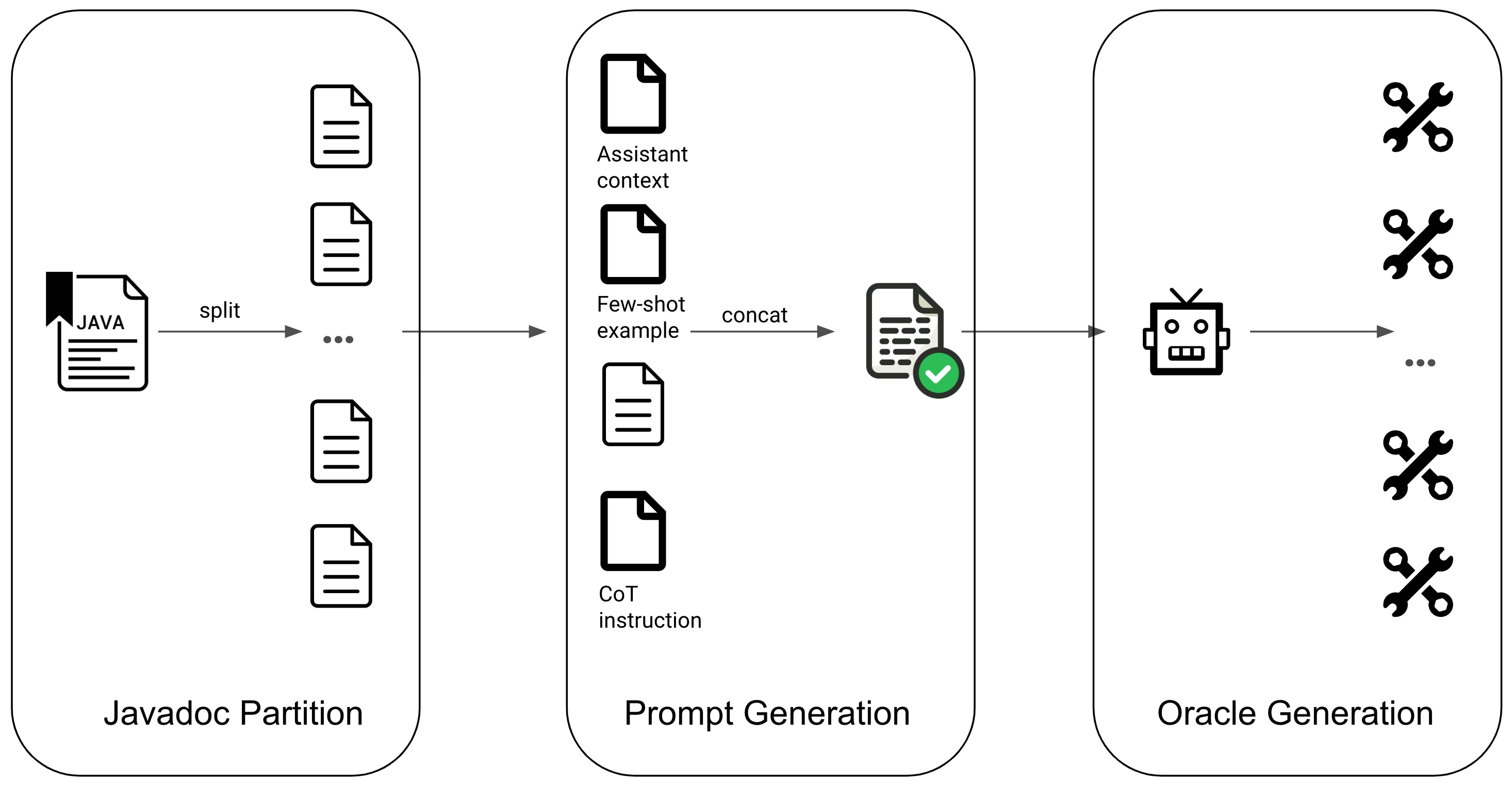}
    \caption{Architecture of Test Oracle Generation}
    \label{arch}
\end{figure}

To tackle the challenge of testing discrepancies between JDK-defined properties and client project implementations, we leverage LLMs as an enabling technology to transform our insights into a technique for automating test oracles.
 In contrast to test oracle generation methods that rely on manually defined oracle templates derived from existing test cases and program source code (e.g., TOGA~\cite{toga}), our approach requires no oracle templates or source code. The only input is natural language documentation, specifically Javadocs.
 We adopt LLMs to directly extract properties in Javadocs and generate useful test oracles for described properties.  Figure \ref{arch} shows the architecture of our work. We introduce some useful prompting techniques to boost the LLMs performance without any fine-tuning. Figure \ref{prompt} shows the structure of our prompt template. The proposed methodology utilizes a structured prompt design to guide the LLMs in generating useful test oracles. Specifically, the prompting process can be divided into several key steps. 

\begin{enumerate}[Step 1:]
\item \textbf{Javadocs Partition}. The original Javadocs is partitioned at the method level to ensure relevant method relationships are captured without overwhelming the LLMs. This prevents the LLMs from forgetting or missing key dependencies between methods, such as the relationship between \textit{equals()} and \textit{hashCode()}.
\item \textbf{Assistant Creation}. The LLMs is framed as a "Software Testing Engineer" to generate test oracles. This persona helps the model focus its reasoning and output on software testing tasks, ensuring it aligns with professional standards.
\item \textbf{Few-shot Learning}. Provide examples that illustrate the desired test oracles for different method properties. By giving exemplar input/output scenarios, the LLMs learns the format and analysis depth required to generate accurate test oracles.
\item \textbf{Chain of Thought Reasoning}. The tasks are broken down into multiple interconnected steps, such as identifying features and generating test oracles. This sequential approach helps the LLMs focus on one sub-task at a time, improving accuracy and coherence.
\end{enumerate}
 The details of prompt engineering are described in the following sections. 

\subsection{Javadocs Partitioning}

Partitioning Javadocs at the method level is essential for improving LLM performance in generating accurate test oracles. Using the entire JDK Javadocs as a prompt poses challenges due to the LLMs' context window limit. A long prompt can cause LLMs to lose track of prior instructions, resulting in incomplete details and potential ``hallucinations'' -- where incorrect test oracles are generated as method relationships blur \cite{an2024make,schwarzschild2024rethinking}. Additionally, LLMs face constraints in both input and output token limits. When presented with unstructured, lengthy prompts, they may inefficiently reiterate method descriptions instead of producing specific, actionable test oracles.

\begin{figure}
    \centering
    \includegraphics[width=\linewidth]{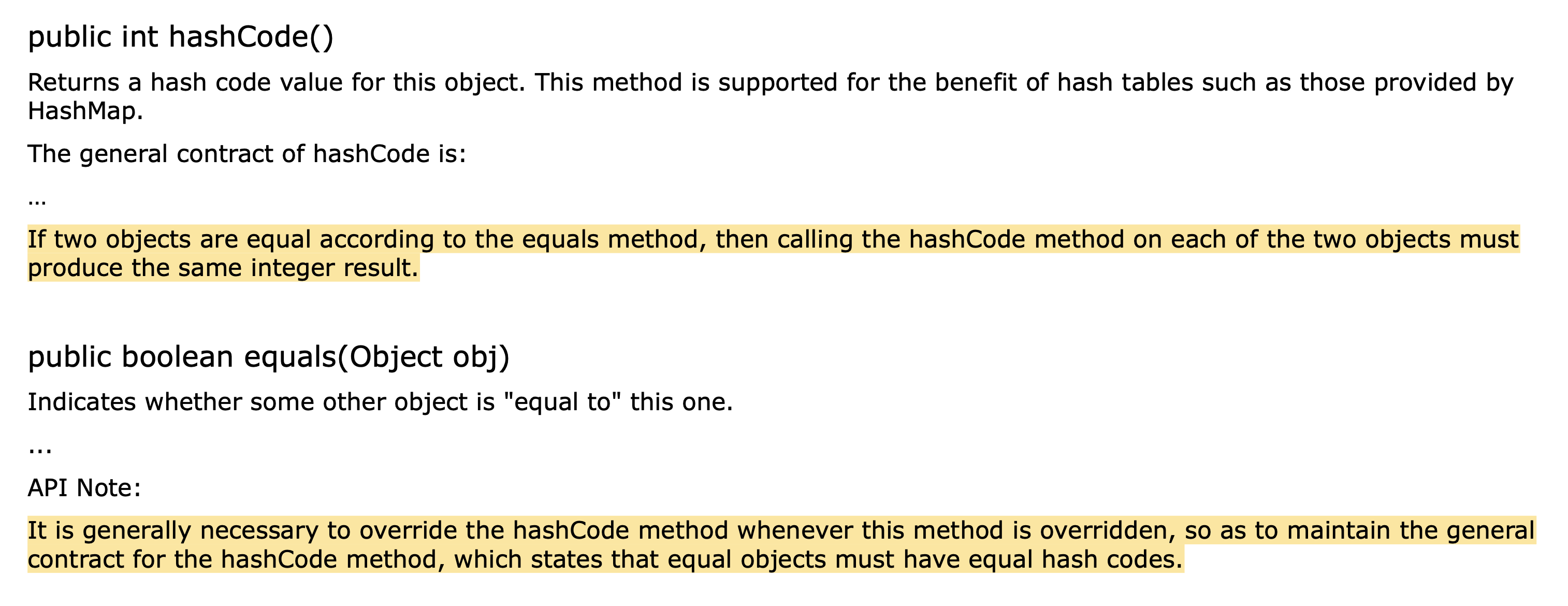}
    \caption{\textit{equals} and \textit{hashCode} method defined in \textit{java.lang.Object}}
    \label{relation}
\end{figure}

Partitioning Javadocs also addresses issues arising from method interdependencies. For instance, in the Java API, methods like \textit{equals()} and \textit{hashcode()} are tightly coupled, as illustrated in Figure \ref{relation}; overriding one usually requires overriding the other. If only one method is included in the prompt, LLMs may overlook these crucial relationships, leading to incomplete or incorrect test oracles. By partitioning at the method level, LLMs can concentrate on specific, related subsets of the Javadocs, ensuring key method dependencies are captured and correctly managed.


We propose a partitioning strategy that segments Javadocs at the method level, grouping related methods, especially those linked through ``See Also'' sections, in the same prompt. This approach helps LLMs maintain relevant context while minimizing the processing load of an entire class or interface. For instance, since method \textit{equals()} references \textit{hashcode()} in its Javadocs, both are included together, enabling the LLMs to grasp their interdependence and generate coherent test oracles, as shown in Figure \ref{consistency}. 
Algorithm 1 details the steps of the Javadocs partitioning process. Each method is treated as a standalone input, with related methods included to provide full context. These method descriptions are then passed as input to the LLMs.

 \begin{figure}
	\centering
	\begin{minipage}[t]{0.5\linewidth}
	\end{minipage}
	\begin{minipage}[t]{0.8\linewidth}
		\begin{lstlisting}[style = Java-github]
boolean checkEqualsHashCodeConsistency(Object x, Object y) {
    if (x != null && y != null && x.equals(y)) {
        return x.hashCode() == y.hashCode();
    }
    return true;
}
            \end{lstlisting}
            \caption{Test oracle for checking \textit{equals()} and \textit{hashCode()} contracts}
    \label{consistency}
	\end{minipage}
 
\end{figure}


\begin{figure}[ht]
  \centering
  \begin{minipage}{.6\linewidth}
\begin{algorithm}[H]
\caption{Javadocs Partition}\label{partition}
\KwIn{Javadocs $j$;}
\KwOut{Descriptions $descriptions$;}
\ForEach{m in j}{
    $description \gets m$
    
    \ForEach{n in m.SeeAlso}{
        \If{n in j}{
            $description \gets description + n$
        }
    }
    $descriptions \gets descriptions + description$
}
\Return{descriptions}
\end{algorithm}
    
  \end{minipage}
\end{figure}

\subsection{Assistant Creation}

We employ assistant creation to guide LLMs in generating test oracles, a method proven effective in prior research~\cite{zhang2023ecoassistant,dong2023towards}. As shown in the <context> section (Line 1-5) of Figure \ref{prompt}, we initiate the prompt by stating, "You are a software testing engineer," which frames the task within software testing. This role assignment directs the LLMs to focus on testing-related reasoning, such as identifying properties described in Javadocs and translating them into executable test oracles, thereby enhancing response relevance and performance.



This approach's effectiveness stems from narrowing the LLM’s broad capabilities to a focused, goal-oriented persona. In practice, this role-based framing influences how the LLM interprets the prompt and prioritizes information. Acting as a software testing engineer, the LLM, when given a Javadocs description, concentrates on key behavioral properties such as preconditions, postconditions, invariants, or expected exceptions -- critical elements for test oracle generation. This ensures that the output not only verifies correctness but also addresses edge cases and potential failure points. Moreover, the assistant creation sets a consistent tone for interaction with the LLM~\cite{gao2024aligning}. Beyond role specification, it reinforces testing principles like accuracy, reliability, and completeness in test oracles. For example, the prompt directs the LLM to ensure coverage of all properties mentioned in the Javadocs, preventing omissions that could lead to incomplete test oracle coverage.


The assistant creation process transforms the LLMs from a general-purpose model into a specialized tool, making it an intermediary between natural language Javadocs and test oracle generation. By defining the assistant’s role early on, we ensure that each subsequent step—parsing method descriptions, identifying test cases, and generating assertions -- is executed with a focused software testing perspective.


\subsection{Few-shot Learning}

Few-shot learning is crucial for improving the LLM's ability to generate high-quality test oracles by providing carefully crafted examples that steer it toward the desired output \cite{ma2023fairness,xu2022prompting}. This technique offers the LLM a small yet representative set of input-output examples, helping it understand the format, depth, and specificity required for effective test oracle generation. This minimizes ambiguity and enables the LLM to generalize from these examples to new, varied contexts.


Our core idea behind few-shot learning is to ``teach'' the LLM how to interpret Javadocs descriptions and translate them into test oracles by presenting a few solved examples. For instance, in the <example> section (Line 7-31) of Figure \ref{prompt}, we provide a Javadocs description for a method like \textit{equals(Object obj)} from the \textit{Object} class and illustrate how to generate a test oracle verifying properties including reflexivity, symmetry, and transitivity. By supplying the model with these well-defined examples, we establish a pattern it can emulate when processing similar method descriptions in new contexts. Few-shot learning also enables the LLM to handle a broad range of method types, from simple methods without dependencies (e.g., \textit{getClass}) to more complex methods with interrelated properties (e.g., \textit{equals} and \textit{hashCode}). 



We also use few-shot learning to enhance the LLM’s ability to handle edge cases, particularly for methods that may exhibit exceptional behavior, such as those that throw exceptions under specific conditions. By providing examples, we can guide the LLM to generate oracles that account for these cases. For instance, we demonstrate how to create a test oracle that verifies a method correctly throws a \textit{NullPointerException} when given a null input, which is an edge case common in Java programs.



\subsection{Chain of Thought}

 We use Chain of Thought, a technique empirically shown to be effective in complex tasks \cite{wei2022chain,chen2023you}, to enhance the LLM’s performance and accuracy in test oracle generation. This method involves breaking down the task into smaller, logically connected steps that guide the LLM through the process in a structured, step-by-step manner. By dividing the task into manageable sub-tasks, prompt chaining helps the LLM stay focused, producing coherent and accurate outputs while avoiding the challenges of handling excessive information at once \cite{li2023structured,diao2023active}.



Our first prompt-chaining step focuses on feature extraction. The LLM is instructed to analyze the Javadocs for a given method and identify all relevant properties or behaviors requiring testing. For example, if the method is \textit{equals(Object obj)}, the LLM is guided to extract properties such as reflexivity, symmetry, transitivity, consistency, and null handling -- each a feature essential for verifying the method’s behavior as specified in the Javadocs.

The second step in the chain is generating test oracles for each identified feature. After successfully extracting the key properties of a method, the LLM is then tasked with creating specific test oracles to verify each property. For example, once the reflexivity of \textit{equals(Object obj)} is identified, the LLM generates a test oracle to confirm that any non-null object equals itself (i.e., \textit{x.equals(x)} should return true).

Finally, we extended prompt chaining to generate advanced test oracles involving exception handling or boundary conditions. For example, after identifying that a method throws a \textit{NullPointerException} when passed a null argument, the next step in the chain prompts the LLM to generate a test oracle confirming that the exception is indeed thrown under these conditions. This step-by-step approach ensures that even complex oracles involving multiple conditions or exceptional behaviors are accurately handled.



This prompt chaining based approach reduces the model’s cognitive load by allowing it to focus on one sub-task at a time, ensuring each step is completed before proceeding. It also improves accuracy, as the LLM is less likely to miss essential features or produce incomplete or incorrect test oracles when working in a sequential, structured manner. Additionally, prompt chaining maintains logical consistency throughout the process by building on the work completed in the previous step, ensuring the test oracles remain relevant to the features identified.

An added advantage of prompt chaining is that it provides implicit feedback at each stage. After feature extraction, the LLM is better equipped to generate meaningful test oracles because it has already developed a clear understanding of the method’s properties. This feedback loop reinforces the model’s focus and helps prevent it from deviating from the task.



 \begin{figure}
	\centering
	\begin{minipage}[t]{\linewidth}
\begin{lstlisting}[style = prompt]
<context>
    You are a software testing engineer. You will be provided with Java method description in <class type 
    here>, and your task is to find all features in method descriptions and generate test oracles for all 
    features one by one. You do not need to generate the whole test cases, just focus on test oracles.
</context>

<examples>
    <description>
        public boolean equals(Object obj) 
        Indicates whether some other object is equal to this one. The equals method implements an equival-
        -ence relation on non-null object references:
            It is reflexive: for any non-null reference value x, x.equals(x) should return true. 
            It is symmetric: for any non-null reference values x and y, x.equals(y) should return true if 
            and only if y.equals(x) returns true. 
            It is transitive: for any non-null reference values x, y, and z, if x.equals(y) returns true and 
            y.equals(z) returns true, then x.equals(z) should return true. 
            It is consistent: for any non-null reference values x and y, multiple invocations of x.equals(y) 
            consistently return true or consistently return false, provided no information used in equals 
            comparisons on the objects is modified. 
            For any non-null reference value x, x.equals(null) should return false. 
        An equivalence relation partitions the elements it operates on into equivalence classes; all the 
        members of an equivale-nce class are equal to each other. Members of an equivalence class are subst-
        -itutable for each other, at least for some purposes.
    </description>
    <oracle>
        For reflexive, the test oracle is:
            boolean checkReflexive(Object x) {
                return x != null ? x.equals(x) : true;
            }
    </oracle>
</examples>

<instruction>
    Use the following step-by-step method to generate test oracles. Remember that you need to generate a 
    test oracle that returns a boolean value rather than an entire test case that can be executed. If 
    necessary, you can use the try catch structure in test oracles to catch exception. Test oracles may 
    require some input, you need to determine the input as well, most time the input should be same as class
    type. No matter in which cases, still return a boolean to indicate whether the feature is satisfied.
        Step 1 - Find all properties and behaviors requiring testing in the Java method description.
        Step 2 - Generate test oracles for each identified feature one by one.
        Step 3 - Generate test oracles for exception handling and boundary conditions.
    This is the Java method description you need to deal with:
    <method description here>
</instruction>
\end{lstlisting}
            \caption{Our prompt template input to LLMs}
    \label{prompt}
	\end{minipage}
 
\end{figure}

\section{Test Oracle Evaluation}
In this section, we evaluate the effectiveness of using Large Language Models (LLMs) for test oracle generation from Javadocs. Our primary goal is to assess the practicality of the generated test oracles and their ability to accurately validate software behavior without relying on project source code. Specifically, we investigate the following research questions:

\textbf{RQ1: Compilability.} How many LLMs-generated test oracles are able to compile successfully?

\textbf{RQ2: Compeleteness.} How complete are the LLMs generated test oracles in covering the key properties described in Javadocs?

\textbf{RQ3: Exception Handling.} How complete are the LLMs generated test oracles in handling exceptions described in Javadocs?

\textbf{RQ4: Quality.} What is the quality of the LLMs generated test oracles? Are they clear and understandable?

\textbf{RQ5: Prompt Ablation.}  How effective are our prompt techniques? What will happen if we drop some of them?
\subsection{Experimental Setup}

To answer the research questions, we designed an experimental setup based on the following key components:

\subsubsection{Large Language Models.} We choose the state-of-the-art GPT-4 model, setting the temperature to 0.7 to balance creativity and consistency in the generated outputs. This configuration has been found to perform well for code generation tasks\cite{endres2024can}, allowing for a diverse but controlled generation of test oracles. We access the GPT-4 model via ChatGPT website.

\subsubsection{Javadoc Dataset.} We choose the most widely used java libraries as our dataset. Specifically, we choose 2 classes from \textit{java.lang} (\textit{Object}, and \textit{String}) and 3 interfaces from  \textit{java.util} (\textit{Map}, \textit{Set}, and \textit{List}). These classes represent a broad spectrum of functionality and are commonly used across Java projects. By using this standardized dataset, we ensure that the generated oracles have strong generalization capabilities, as they can be applied across various Java projects that depend on these Java libraries.


\subsection{RQ1: Compilability}


The first research question in evaluating the generated test oracles is their compilability, as non-compiling oracles cannot be used in testing. We compile the test oracle methods in a standard Java environment and manually review errors for non-compiling oracles to assess if they still represent the intended properties in the Javadocs. Such oracles are considered correct. Due to similar overload functions in the JDK Javadocs, LLMs may generate oracles with identical names and parameters, causing compilation errors. In these cases, we only adjust function names, preserving parameters and oracle content. We measure the percentage of oracles that compile without further modification.

\begin{table}[]
\begin{tabular}{lrrrr}
\toprule
  Class & \#Methods  & \#Oracles & \#Compilable Oracles  & \#Correct Oracles \\
  \hline
  \textit{java.lang.Object}  & 11 & 38 & 36(94.7\%) & 38(100\%)\\
  \textit{java.lang.String}  & 77  & 158 &  151(96.7\%)&  155(98.1\%)\\
  \textit{java.util.Set} &  20 &  54 &  53(98.2\%) & 53(98.2\%)  \\
  \textit{java.util.List} & 32 & 87 & 86(98.9\%) & 86(98.9\%)\\
  \textit{java.util.Map} & 25 & 88 & 87(98.9\%) & 88(100\%)\\
  Total & 165 & 428 & 418(97.7\%) & 423(98.8\%)\\
  \bottomrule
\end{tabular}
  \caption{RQ1: Evaluation of Compilability}
  \label{compile}
\end{table}


Table \ref{compile} presents the results of RQ1, with columns for the number of methods, generated oracles, compilable oracles, and correct oracles. We observe that: (1) Over 97\% of test oracles generated by LLMs compile successfully, and (2) 423 out of 428 oracles (98.8\%) are correct. Correctness differs from compilability, as some test oracles require helper functions or hypothetical classes. For instance, the test oracle in Figure \ref{hypo} targets the \textit{Object.clone()} method by checking if the cloned object remains independent of the original via field changes. Since  \textit{Object} lacks mutable fields, LLMs use a hypothetical class with fields for simulation, making the oracle correct but unable to compile due to the undefined class. Such oracles, though non-compilable, are easy to fix and provide a useful guideline for test oracle development.


Incorrect oracles fall into two categories: (1) Type errors, such as \textit{int actual = set.stream().count();}, where the return type of \textit{count()} is \textit{long}, and implicit casting from \textit{long} to \textit{int} is not allowed; (2) Syntax errors, such as \textit{actualResult == mainStr.indexOf(subSeq.toString()) != -1;}, where the intended behavior is to compare \textit{actualResult} with the result of \textit{mainStr.indexOf(subSeq.toString()) != -1}. However, due to incorrect execution order, it first compares \textit{actualResult} with \textit{mainStr.indexOf(subSeq.toString())}. Adding parentheses corrects this to make a valid oracle.

\begin{findingbox}{Answer to RQ1: Compilability of Test Oracles}
Overall, our approach effectively generates compilable test oracles (97\%) without additional compilation tools. Even non-compilable or incorrect oracles offer valuable guidance for manual test oracle development, as they contain only minor, easily correctable errors.
\end{findingbox}

 \begin{figure}
	\centering
	\begin{minipage}[t]{0.5\linewidth}
	\end{minipage}
	\begin{minipage}[t]{0.8\linewidth}
		\begin{lstlisting}[style = Java-github]
boolean checkCloneIndependency(Object x) throws CloneNotSupportedException {
    Object original = x.clone();
    Object clone = original.clone();
        
    // Assuming clone modifies a mutable field as a simple example
    if (original instanceof CloneExample) { // CloneExample is a hypothetical class with mutable fields
        ((CloneExample) clone).setMutableField(new Object());
    }

    return !clone.equals(original);
}
            \end{lstlisting}
            \caption{An example of correct but not compilable oracle}
    \label{hypo}
	\end{minipage}
 
\end{figure}

\subsection{RQ2: Completeness}

This research question assesses whether generated test oracles cover all intended properties of each method as defined in the JDK Javadocs, including all assertion and exception cases. Note that exception oracles and assertion oracles may overlap, as LLMs can generate effective oracles that simultaneously verify functionality and expected exceptions, as shown in Figure \ref{exc}. Table \ref{assert} provides the overall results: \textit{\#Documented} indicates the properties identified in Javadocs, \textit{\#Generated} represents the properties covered by generated oracles, and \textit{\#Checked} shows the properties successfully checked by test oracles. Precision is calculated as \textit{\#Checked / \#Generated}, and Recall as \textit{\#Generated / \#Documented}.


We also examine how LLMs manage varying levels of complexity in generating assertion oracles.

\begin{figure}
	\centering
	\begin{minipage}[t]{0.44\linewidth}
		\begin{lstlisting}[style = Java-github]
/**
 * Test oracle for checking if isEmpty correctly identifies an empty list.
 *
 * @param list the list to check
 * @return true if isEmpty returns true for an empty list and false for a
 *         non-empty list, false otherwise
 */
boolean checkIsEmpty(List<?> list) {
    boolean empty = list.isEmpty();
    if (list.size() == 0) {
        return empty; // Should be true if the list is indeed empty
    } else {
        return !empty; // Should be false if the list is not empty
    }
}

/**
 * Test oracle for checking if remove(Object o) correctly removes the first occurrence of the element.
 *
 * @param list the list to be checked
 * @param o    the element to be removed
 * @return true if the element is correctly removed and method return true false otherwise
 */
<E> boolean checkElementRemoval(List<E> list, E o) {
    int originalSize = list.size();
    boolean contains = list.contains(o);
    boolean result = list.remove(o);
    boolean newSizeCorrect = list.size() == (contains ? originalSize - 1 : originalSize);
    return result == contains && newSizeCorrect;
}
            \end{lstlisting}
            \caption{An example of implicit conditions}
            \label{implicit}
	\end{minipage}
 \hspace{.3in}
	\begin{minipage}[t]{0.44\linewidth}
		\begin{lstlisting}[style = Java-github]
// Oracle to verify that the wait is indefinite without notify
boolean checkIndefiniteWait(Object obj) {
    Thread notifyingThread = new Thread(() -> {
        try {
            Thread.sleep(100); // Delay to ensure main thread is waiting
            synchronized (obj) {
                obj.notify();
            }
        } catch (InterruptedException e) {
            Thread.currentThread().interrupt();
        }
    });
    
    long startTime = System.currentTimeMillis();
    synchronized (obj) {
        try {
            notifyingThread.start();
            obj.wait(); // This should wait until it is notified above
            long waitTime = System.currentTimeMillis() - startTime;
            return waitTime >= 100 && waitTime < 200; // Check that wait was indeed waiting until notified
        } catch (InterruptedException e) {
            return false; // If interrupted, not handling as indefinite wait
        }
    }
}

            \end{lstlisting}
            \caption{An example of complex conditions}
            \label{complex}
	\end{minipage}
\end{figure}

\begin{table}[]
\begin{tabular}{lrrrrr}
\toprule
  Class & \#Documented  & \#Generated & \#Checked & Precision(\%) & Recall(\%) \\
  \hline
  \textit{java.lang.Object}  & 33  & 30 & 29 & 96.7 & 90.9\\
  \textit{java.lang.String}  & 162 & 151 & 144 & 95.4 & 93.2 \\
  \textit{java.util.Set} &  44 & 39  & 36  & 92.3 & 88.6\\
  \textit{java.util.List} & 71 &  63 & 62 & 98.4 & 88.7 \\
  \textit{java.util.Map} & 80 & 69 & 67 & 97.1  & 86.3\\
  Total & 390 & 352 & 338 & 96.0 & 90.3\\
  
  \bottomrule
\end{tabular}
  \caption{RQ2: Evaluation of Assertion Oracles}
  \label{assert}
\end{table}

(1) \textbf{Explicitly defined properties.} LLMs can generate simple assertion oracles directly from property descriptions in Javadocs (e.g., Figure \ref{sym}). In our experiments, we found no instances where LLMs ignored explicitly defined properties. Here, LLMs leverage their training data to utilize JDK helper functions without needing additional prompt context.


(2) \textbf{Implicit conditions.}
For instance, in the \textit{java.util.List} interface, the \textit{isEmpty()} method is described as ``Returns true if this list contains no elements.'' As shown in Figure \ref{implicit}, although there is no explicit description of the \textit{size()} method, LLMs can infer that \textit{isEmpty()} should yield the same boolean result as \textit{size() == 0}. Leveraging a large training dataset, LLMs can call functions not directly mentioned and create a structure that accurately represents the intended Javadocs behavior. Remarkably, LLMs can generate oracles for implicit details, like reproducibility. For example, although reproducibility is not specified for the \textit{getClass()} method in \textit{Object} Javadocs, LLMs produced an oracle that repeatedly calls this method, verifying consistent results. This capacity to link properties beyond explicit documentation demonstrates the model’s nuanced understanding from training.


(3) \textbf{Complex conditions.}
LLMs can simulate complex scenarios, such as multithreading (e.g., Figure \ref{complex}), and detect correlations between methods. For interdependent methods like \textit{equals()} and \textit{hashcode()}, we evaluate whether generated oracles maintain consistency across methods, covering both individual behaviors and their interrelationships. The generated oracles respect these dependencies, demonstrating the model’s capability to uphold consistency across related functions, as shown in Figure \ref{consistency}.

We observe that LLMs can generate test oracles for most properties in JDK Javadocs with high precision. Their effectiveness is due to extensive training data, which improves their ability to identify Javadoc features. Even when Javadocs lack details for a fully correct oracle, LLMs can suggest sample oracles that guide property verification. For instance, in \textit{Map} subclasses where order matters, LLMs can use ordered sets like \textit{TreeSet} to demonstrate ordering. These oracles offer developers helpful guidance.

\begin{findingbox}{Answer to RQ2: Completeness of Assertion Oracles}
The results demonstrate LLMs' effectiveness in generating assertion oracles that are both syntactically correct and semantically accurate, producing broadly applicable oracles. Overall, LLMs cover 90.3\% of the properties identified in JDK Javadocs, with 96.0\% of the generated oracles successfully checking these properties.
\end{findingbox}

\subsection{RQ3: Exception Handling}
Exception handling is essential in testing, and this research question examines whether the generated test oracles correctly check for expected exceptions under specified conditions. Our exception oracles return a boolean value to indicate correct exception handling. Specifically, the test oracles employ a \textit{try-catch} structure to catch and filter for expected exceptions, returning false for any unexpected ones. Table \ref{exception} shows the results of exception oracles, with columns defined as in Table \ref{assert}.

\begin{figure}
	\centering
	\begin{minipage}[t]{0.44\linewidth}
		\begin{lstlisting}[style = Java-github]
/**
 * Test oracle to check if codePointAt method correctly handles index validation.
 * 
 * @param str   the string to test
 * @param index the index of the code point to retrieve
 * @return true if the method correctly throws IndexOutOfBoundsException when necessary, false otherwise
 */
boolean checkIndexValidation(String str, int index) {
    try {
        int result = str.codePointAt(index);
        return true; // No exception means index is within valid range.
    } catch (IndexOutOfBoundsException e) {
        return index < 0 || index >= str.length();
    } catch (Exception e) {
        return false; // Handle unexpected exceptions.
    }
}
            \end{lstlisting}
	\end{minipage}
 \hspace{.3in}
	\begin{minipage}[t]{0.44\linewidth}
            \begin{lstlisting}[style = Java-github]
boolean checkConcurrentModificationException(Map map, BiConsumer<? super K, ? super V> action) {
    try {
        Iterator<Map.Entry<K, V>> it = map.entrySet().iterator();
        if (it.hasNext()) {
            map.remove(it.next().getKey()); // Modify map during iteration
        }
        map.forEach(action); // Attempt to perform action after modification
        return false; // If it reaches here, no ConcurrentModificationException was thrown
    } catch (ConcurrentModificationException e) {
        return true; // Correct behavior, exception was thrown
    }
}


            \end{lstlisting}
	\end{minipage}
 \caption{An Example of generated exception oracles}
            \label{exc}
\end{figure}


The left side of Figure \ref{exc} shows an example of an exception oracle, where we use a \textit{try-catch} structure to check if the \textit{codePointAt(int index)} method correctly handles out-of-bounds indices. If an \textit{IndexOutOfBoundsException} is thrown, the oracle verifies the index validity and returns the result, ensuring it can detect abnormal behavior. On the right is an incomplete exception oracle: a \textit{ConcurrentModificationException} should be thrown when a modification occurs during iteration or if multiple threads modify the \textit{Set} concurrently. This oracle simulates only the first case, failing to cover concurrent modifications, so it incompletely checks the exception. However, the oracle itself remains functionally correct. Notably, even without \textit{ConcurrentModificationException} being explicitly mentioned in the Javadocs, LLMs infer the exception through the \textit{fail-fast} property. Although LLMs may lack full context for certain exceptions, their ability to generate even incomplete oracles demonstrates impressive potential and highlights their strength in managing complex, context-limited scenarios.

\begin{table}[]
\begin{tabular}{lrrrrr}
\toprule
  Class & \#Documented  & \#Generated & \#Checked & Precision(\%) & Recall(\%) \\
  \hline
  \textit{java.lang.Object}  & 11 & 11 & 11 & 100 & 100\\
  \textit{java.lang.String}  & 28 & 28 & 27 & 96.4 & 100\\
  \textit{java.util.Set} & 24  & 23  & 22  & 95.7 & 95.8 \\
  \textit{java.util.List} & 62 & 62  & 61 & 98.4 & 100 \\
  \textit{java.util.Map} & 57 & 56 & 54 &  96.4 & 98.2\\
  Total & 182 & 180 & 175 & 97.2 & 98.9 \\
  \bottomrule
\end{tabular}
  \caption{RQ3: Evaluation of Exception Oracles}
  \label{exception}
\end{table}

\begin{findingbox}{Answer to RQ3: Completeness of Exception Oracles}
The results demonstrate LLMs' effectiveness in generating exception oracles that are both comprehensive and accurate. Overall, LLMs can generate oracles for 98.9\% exceptions defined in JDK Javadocs, and 97.2\% generated oracles can catch corresponding exceptions correctly.
\end{findingbox}

\subsection{RQ4: Oracle Quality and Understandability}

 A critical aspect of high-quality test oracles is adherence to software engineering best practices, including meaningful variable naming, logical clarity, and coding standards, which enhance comprehensibility and maintainability for developers. This research question evaluated the generated oracles focused on several key qualities:

\subsubsection{Naming and Documentation.}
The structured prompts used in this study consistently guided LLMs to create functionally and semantically appropriate names for test oracles based on the properties being verified. For instance, in \textit{checkElementRemoval} (Figure \ref{implicit}), when complex properties with multiple sub-properties were identified, LLMs effectively used distinct, meaningful variable names for each sub-property, enhancing clarity and maintainability.

\subsubsection{Comments and Javadocs.}
LLMs demonstrated the capability to integrate relevant comments within test oracles, clarifying the properties under test. Additionally, they generated suitable Javadoc comments for each test oracle, as seen in Figure \ref{implicit}, detailing tested properties along with expected inputs and outputs, thus ensuring clarity in understanding and usage.

\subsubsection{Code Clarity and Structure.}
The generated test oracles showed clear and concise code structure, following unit testing best practices. LLMs effectively used ternary expressions to encapsulate property checks and structured exception handling to capture potential runtime anomalies. Where feasible, LLMs also combined handling of two exceptions within a single oracle, ensuring correctness while improving oracle effectiveness.

\begin{findingbox}{Answer to RQ4: Oracle Quality and Understandability}
Overall, LLMs generate oracles with human-readable variable names and a clear code structure, making them easy to understand and maintain.
\end{findingbox}

\subsection{RQ5: Prompt Ablation}
In designing our prompt, we experimented with various techniques and ultimately selected the aforementioned four. This research question evaluates the impact of specific prompt engineering strategies on the effectiveness and reliability of test oracle generation.





\subsubsection{Assistant Creation.}
Excluding the assistant creation phase from our prompts significantly reduced the effectiveness of the generated oracles. In some instances, LLMs shifted from producing executable code to merely outlining test strategies in a descriptive manner.

\subsubsection{Few-shot Learning.}
Without few-shot learning, LLMs struggled to grasp task requirements, often defaulting to test cases with hard-coded values rather than adaptable test oracles. This limitation was especially evident in features needing generalization, such as those in the \textit{Map} class.

\subsubsection{Chain of Thought (CoT).}
The absence of CoT in prompts led to ``lazy'' responses, where not all Javadocs features were identified or were incorrectly merged into a single test oracle. This compromised effectiveness, especially when distinguishing between exceptions like \textit{IllegalArgumentException} and \textit{IllegalMonitorStateException}.

\subsubsection{Javadocs Partitioning.}
Javadocs partitioning proved essential for managing LLM output token limitations, ensuring that LLMs did not merely describe features but generated actionable test oracles for specific methods.

\begin{findingbox}{Answer to RQ5: Effectiveness of Prompt Techniques}
All of our prompt techniques are essential for enabling the LLMs to understand the task and generate accurate test oracles. Removing any of these techniques causes the LLMs to suffer from ablation, leading it to generate descriptions instead of oracles or simply repeat information from the prompts.
\end{findingbox}

\section{Threats to Validity}

\textbf{External Threats.} 
The primary external threats to our approach stem from the inherent randomness of LLMs. This randomness is unavoidable and also contributes to the diversity of generated program variants and inputs. Our approach has certain limitations tied to the use of LLMs, including potential obsolescence due to the rapid evolution of AI technology. 
Despite potential variations, the LLMs demonstrated consistent performance in oracle generation, with test oracle effectiveness unaffected by differing comment styles. To mitigate this threat and enhance the representativeness of our results, we conducted experiments on widely used JDK classes from \textit{java.lang} and \textit{java.util}.

\textbf{Internal Threats.} 
Our evaluation relies on manual verification of test oracle correctness, introducing a subjective element into our analysis. To migrate this, two authors independently performed the checks. Any discrepancies in our inspection results were further checked to ensure the correctness of the manual inspection. Besides, the ``black box'' nature of ChatGPT and similar LLMs can obscure the rationale behind model outputs, making it challenging to interpret or replicate results precisely. To mitigate this, we use a Chain of Thought prompting approach, enhancing transparency and making it easier to trace from Javadocs inputs to the generated oracles. 

\section{Related work}
\textbf{Test Oracle Generation} is a critical component of automated software testing, as test oracles define the intended behavior of a software system. Numerous studies have focused on advancing methods for test oracle generation. Overall, these methods can be divided into two categories: specification mining method and neural method\cite{fontes2021using,kamaraj2023hybridized}. Specification mining methods rely on a restricted format of documentation and a set of handcrafted rules to infer exceptions and assertions. @TComment\cite{tan2012tcomment} defines natural language patterns along with heuristics to infer nullness properties. More recently, MeMo\cite{blasi2021memo} uses equivalence phrases in Javadocs comments to infer metamorphic relations (e.g., \textit{sum(x,y) == sum(y,x)}). However, if developers do not adhere to a standardized documentation format or omit documentation entirely, these methods typically fail to extract meaningful oracles from most real-world software components. Recently, neural models have also been used to generate test oracles. NLP-based methods, such as ATLAS\cite{watson2020learning} and AthenaTest\cite{tufano2020unit} have been shown to outperform the specification mining methods. However, these methods rely solely on the implementation and the System Under Test (SUT), overlooking valuable information in documentation. This poses a problem: if the implementation is incorrect and intended behavior is inferred from the code itself, testing effectiveness is compromised. More recently TOGA\cite{toga} established a new state of the art by a large margin. TOGA is not a generative model; it constrains the search space by manually defining oracle templates. For each assertion, it uses a ranking model to select the most relevant template. This approach, however, depends heavily on predefined templates and struggles to cover all possible scenarios.

\textbf{Large Language Models (LLMs)} show effectiveness for various software development tasks recently, including program synthesis\cite{jain2022jigsaw} and test generation\cite{xia2024fuzz4all}. LLMs can generate program code\cite{jain2022jigsaw}\cite{ugare2024improving}\cite{spiess2024quality} from natural language prompts, by associating documentation text with code from a large training set\cite{wang2023large}. TiCoder\cite{lahiri2022interactive} adopts LLMs to generate the whole test cases to formalize the user intent, which include prefix and oracle. TOGLL\cite{hossain2024togll} fine-tuned LLMs to generate test oracles from project implementations and documentations,
%
ensuring that implementations align with their own documentation. However, fine-tuning demands high-performance hardware, making it expensive for developers with limited resources.
The main difference between these methods and ours is that we do not rely on predefined rules or Java code. Our approach requires only JDK Javadocs as input, making our oracles applicable across any Java project. Our goal is to generate general test oracles from the JDK, ensuring conformance to universal rules (e.g., JDK contracts) rather than project-specific constraints. To our knowledge, we are the first to leverage LLMs to address this problem.

\section{Conclusions}
In this paper, we introduced a novel approach for test oracle generation from JDK Javadocs using LLMs. Our method focuses on generating test oracles for clients of widely used Java libraries, e.g., \textit{java.lang} and \textit{java.util} packages. The key insight is that Javadocs that provide a rich source of information can enable automated generation of test oracles without any code implementation. We use large
language models as an enabling technology to embody our insight into a framework for test oracle
automation, and propose a prompt template which is flexible and allows generation of test oracles for a variety of JDK Javadocs. Our approach offers a promising method to validate that clients must conform to the subtle and easy to overlook Java language contracts given in JDK Javadocs rather than only adhering to their own project requirements. The experimental results show that LLMs can generate highly applicable and comprehensive test oracles, capable of accurately expressing expected behaviors and intended exceptions in JDK. 



\bibliographystyle{ACM-Reference-Format}
\bibliography{main}

\end{document}